\documentclass[prb,amsmath,amssymb,twocolumn,letterpaper]{revtex4}
\usepackage{graphicx}

\begin{document}

\title{Low frequency noise in ballistic Nb/Ag point contacts}

\author{Jian Wei, Goutam Sheet, and Venkat Chandrasekhar}
\affiliation{Department of Physics and Astronomy, Northwestern University, Evanston, IL 60208, USA}

\date{\today}

\begin{abstract}
We report on the low frequency noise in the ballistic point-contacts between a silver tip and a niobium foil. The ballistic nature of the point-contacts is confirmed by Andreev reflection spectroscopy at low bias voltage with the Nb foil cooled below its superconducting transition temperature ($T_c$). We find that the voltage dependence of the low frequency noise differs for different point-contacts with varying contact resistances. At high bias voltages, random two level fluctuations appear and dominate over the background $1/f$ noise. From analysis of the Andreev reflection spectra, we show that such two level fluctuators may give rise to depairing of the superconducting order parameter. 
\end{abstract}
\maketitle

The transport properties of a point-contact (PC) between two metals are determined by whether the PC is in the ballistic, diffusive, or thermal regime.~\cite{Naidyuk2005} In the ballistic regime, the size of the constriction $d$ is significantly smaller than the mean free path $l$, and the resistance is due to the limited number of electrons available in the cross-section area of the PC, i.e., the  Sharvin resistance $R_{Sh}=16\rho l/3\pi d^{2}$. In the thermal regime (where $d$ is larger than $l$), the point-contact resistance  is mainly given by the  Maxwell resistance $R_{M}=\rho/d$.  
When one of the metals forming the PC becomes superconducting, $R_{Sh}$ can decrease by up to a factor of two due to Andreev reflection at the point-contact. In this case, the bias-dependence of the point-contact conductance can be analyzed by a theory developed by Blonder, Tinkham and Klapwijk (BTK).~\cite{Blonder1982} Such a bias-dependence, the so-called point-contact Andreev reflection (PCAR) spectrum, has been successfully applied to probe the superconducting order parameters in conventional~\cite{Artemenko1979} and unconventional~\cite{Deutscher1999,Chen2008nature} superconductors. It is believed that measurement of the noise in PCs formed on superconductors might provide more information than simple current-voltage characteristics (IVC) measurements,~\cite{Blanter2000} e.g., about the pairing mechanism of the unconventional superconductors.~\cite{Lofwander2003,Caplan2010prl} Here we study the noise in ballistic PCs between Ag and Nb, a conventional superconductor with $T_c \sim 9.2$ K.

\begin{figure}
\includegraphics[width=8cm]{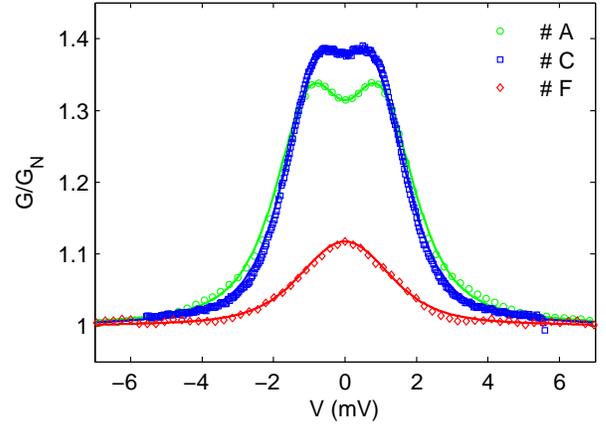}
\caption  {\small Bias dependence of normalized conductance of several PCs at 5.8 K. The solid lines are fits with the BTK model. The fitting parameters are listed in Table 1.} 
\label{fig_dVdI}
\end{figure}

\begin{table}
\caption{Fitting parameters of three PCs: $R_{N}$ is the normal state resistance; $d$ is the estimated diameter of the PC from $R_{N}$; $Z$ is the barrier strength in BTK theory; $\Delta$ is the superconductive gap; $\Gamma$ is the depairing parameter.}
 \begin{tabular}{cccccc}
 \hline\hline
No. & $R_{N}$ &$d$ & $Z$  & $\Delta$ & $\Gamma$    \\
  & ($\Omega$) & (nm) &   &  (mV) & (mV) \\
\hline
A & 14.7 & 9.87 & 0.5 & 1.07 & 0.04    \\
C & 4.72 & 17.4 & 0.47 &  0.94 & 0  \\
F & 35.4 & 6.36 & 0.73 &  0.61 & 0.5  \\
\hline\hline
\end{tabular} 
\end{table}

The measurements are done in a liquid He storage dewar. The sample is mounted on a solid disk of copper (sample holder) installed at the end of a low-temperature scanning probe microscope. There is a heater and a thermometer attached to the sample holder in order to vary the sample temperature. In order to form the ballistic point-contacts, a sharp tip of Ag is brought in contact with the sample by a sophisticated piezo-driven coarse approach mechanism that is installed in the low-temperature scanning probe microscope. The quality of the point-contact is judged by the nature of spectra and comparing it with known spectral features. 

Figure 1 shows the normalized conductance and fits using BTK theory for three PCs that were used for systematic measurement of the bias dependent noise. The fitting parameters are listed in Table 1. The diameter of the ballistic PCs labeled as `A', `C', and `F' respectively are estimated  following the expression for $R_{Sh}$ which leads to $d\approx 37.9/\sqrt{R_{N}}$ (nm), where $R_{N}$ is the normal state resistance at zero bias.~\cite{Naidyuk2005} For all three PCs, the estimated $d$ is smaller than the mean free path of bulk silver ($\sim$ 100 nm) at liquid helium temperature and therefore satisfy the criteria for ballistic PCs. The superconducting gap $\Delta$ obtained from the fitting is lower than that of bulk Nb at the bath temperature ($\sim$ 1.2 meV). This may be attributed to the presence of a natural oxide layer on the niobium foil.~\cite{Halbritter1987}  The fitted depairing parameter $\Gamma$ for PC `F' is significantly higher than that for the other two PCs. This is unusual for a simple weak-coupling conventional BCS superconductor. In the high temperature cuprate superconductors, a large $\Gamma$ is found  which is related to the oxygen stoichiometry at the surface~.\cite{Plecenik1994prb} We note that it is possible to fit the normalized conductance with an elevated temperature instead of using a large $\Gamma$. However, the temperature of the point-contact that is in the ballistic regime should not be enhanced over the bulk as there is statistically no inelastic scattering and hence, no dissipation in the point-contact.  The high contact resistance and consequent small contact size, the absence of conductance dips associated with Maxwell's resistance,~\cite{Sheet2004prb} and the temperature independent resistance above $T_{C}$, all suggest that the contact `F' remains in the ballistic regime.

\begin{figure}
\includegraphics[width=8cm]{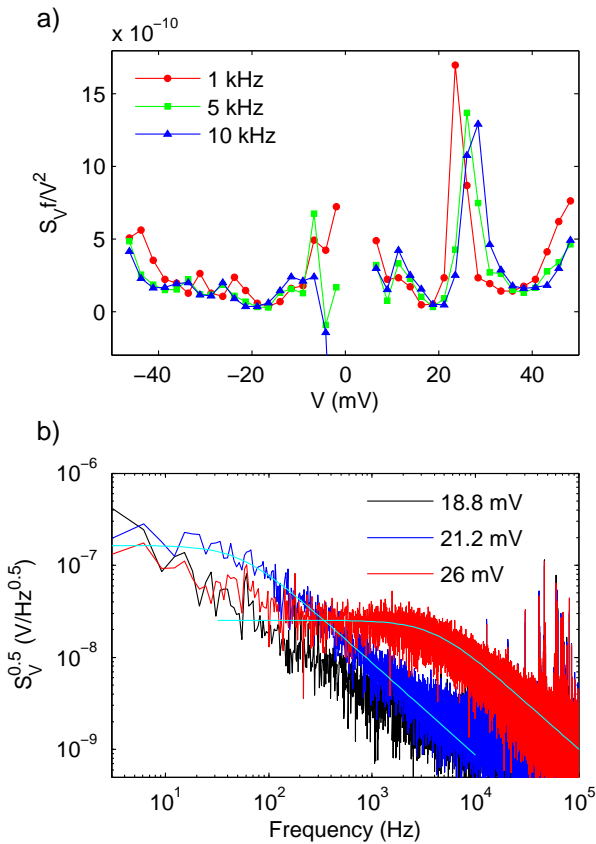}
\caption  {\small Voltage dependent noise of PC `A' at 5.8 K. a): Bias dependence of the normalized voltage noise power. Near zero bias data are not shown since the background noise would lead to artificial effect after divided by $V^{2}$. b): Noise power spectrum density ($\sqrt{S_{V}}$) at several bias voltages corresponding to the peak in a). The cyan lines are fits using Eq.~(1), showing clearly the frequency knee associated with a single TLF.} 
\end{figure}

Now we concentrate on the results of the noise measurements. It is well known~\cite{Ralls1991,Holweg1992prb} that when the size of the point contacts shrinks to nanometer size, the low frequency noise at some finite bias can be dominated by a single two-level fluctuations (TLF), which means the resistance switches between two discrete resistance levels and the corresponding power spectrum is a Lorentzian given by~\cite{Machlup1954,Ralls1991,Holweg1992prb}
\begin{equation}
 S_{R}(f)=\frac{S_{0}\tau_{eff}}{1+(2\pi f \tau_{eff})^{2}},
\end{equation}
where $S_{0}$ is the integrated power and $\tau_{eff}$ the effective time (see e.g., Fig.~2b for the power spectrum and inset of Fig.~4b for the time trace). Here we focus on the difference of the noise spectra for the three ballistic PCs characterized by PCAR. To measure the voltage dependent noise, a DC bias is applied with a current source realized by putting a 3 k$\Omega$ metal film resistor (much larger than the PC resistance) in series with a filtered voltage source. The voltage fluctuation signal from the PC is AC coupled to home-made battery-powered low noise amplifiers, and the output signals after amplification are sent to data acquisition cards and are analyzed by a computer after digitization. In the frequency window of our experimental set-up (up to 100 kHz), the noise spectrum is either of a $1/f$ type or a combination of a dominating TLF and $1/f$ noise background, as shown in Fig.~2, 3, and 4.

\begin{figure}
\includegraphics[width=8cm]{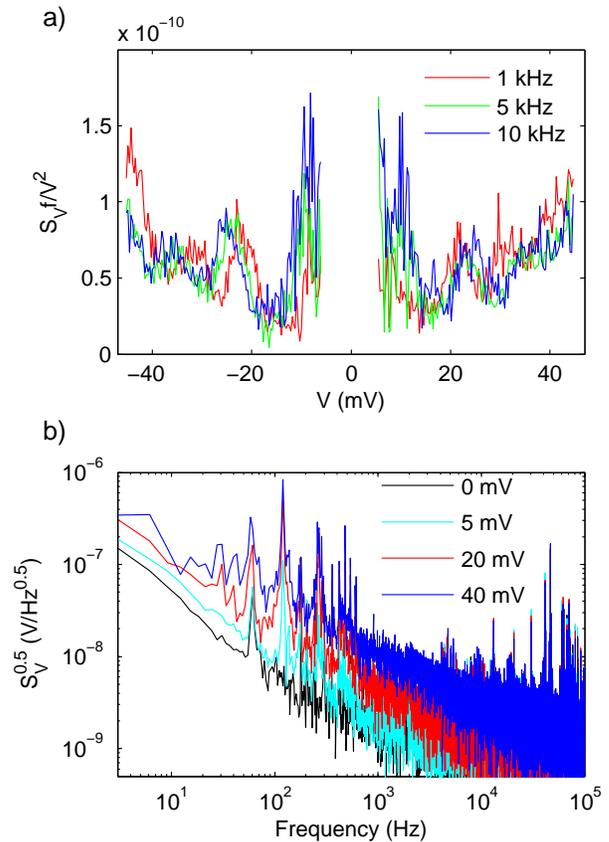}
\caption  {\small Voltage dependent noise of PC `C' at 5.8 K. a): Bias dependence of the normalized voltage noise power. b): Noise power spectrum density ($\sqrt{S_{V}}$) at several bias voltages. } 
\end{figure}

The empirical expression for $1/f$ noise is~\cite{Hooge1969,Weissman1988} $S_{R}(f)/R^{2}=\alpha/N_{c}f$, where $N_{c}$ is the number of charge carriers, and $\alpha$ is a sample-specific constant. To compare the measured noise power with the $1/f$ line shape for different point contacts, we normalize the measured voltage noise power $S_{V}$ with the frequency $f$ and the bias voltage $V$ ($S_{V}/V^{2}=S_{R}/R^{2}$ for near Ohmic region) to obtain a measure of $\alpha/N_c$. In Fig.~2a, $S_{V}f/V^{2}$ at 1, 5, 10 kHz are shown for PC `A'. The scaling of frequency is reasonably good except at around -8 mV and +25 mV, which confirms $1/f$ type spectra in most bias voltage ranges. The magnitude of $fS_{V}/V^{2}$ of the $1/f$ background is of the order of $1\times 10^{-10}$, consistent with previous results for large metal nanobridges.~\cite{Ralls1991} 

Between 20 mV and 30 mV, there are strong peaks with position shifted for different frequencies, which are due to the presence of discrete TLF's and are demonstrated clearly by the Lorentzian line shape spectra in Fig.~2b. The shift of the knee to higher frequency for larger bias was previously observed for ballistic noble-metal nanobridges and was attributed to defect heating.~\cite{Ralls1991,Holweg1992prb}  Such TLF's were also suggested to explain the noise-voltage dependence for shear type PCs,~\cite{Akimenko1984,Matej1993} but has not been investigated for the needle-anvil type heterocontacts employed in this work.  Using a simplified equation $\tau_{eff}=\tau_{0}\exp(\epsilon/V)$, from fits in Fig.~2 we find the attempt frequency $\tau_{0}\approx10^{-12}$ S, and the activation energy $\epsilon \approx$ 483 mV, consistent with previous results for nanobridges.~\cite{Holweg1992prb} We note that TLF's are not universal for PCs. As was shown recently, TLF's are absent in tunable break junctions (another type of PC), and  scaling of the $1/f$ noise with resistance was found from the ballistic to the diffusive regime.~\cite{Wu2008prb}

\begin{figure}
\includegraphics[width=8cm]{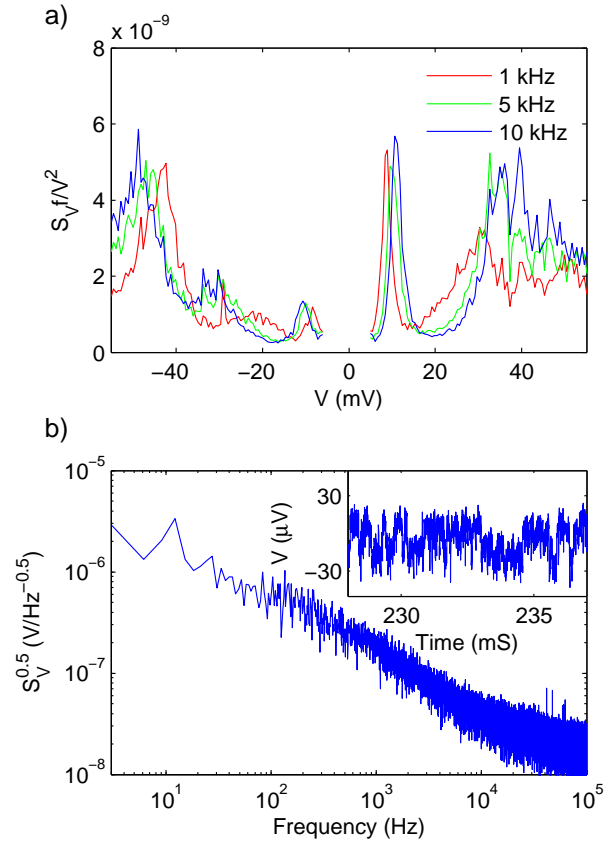}
\caption  {\small Voltage dependent noise of PC `F'. a): Bias dependence of the normalized voltage noise power at 5.8 K. b): A typical noise power spectrum  with 1/f background and a higher spectral weight at around $10^{2}-10^{3}$ Hz due to TLF's, measured at $V\sim$80 mV and at 23 K. The inset shows a small portion of the time trace, where the two main discrete levels and the typical switching time about a few mS are visible.} 
\end{figure}

For PC `C', the influence of TLF's on the noise power spectra is smaller than that of PC `A'. As shown in Fig.~3, there are peaks around $\pm$10 mV, $\pm$25 mV. Although the shift of peak position at different frequencies is still obvious, the magnitude of these peaks are much smaller than that of PC `A'. We repeated the noise  measurements at 5.8 K and 13 K and found that the change of peak position and peak magnitude is small for the peaks around $\pm$10 mV and other peaks in the negative bias side, but large for the higher positive bias side. We also note that there is no direct correlation between the position of TLF's and position of the phonon peaks (around 15 mV) derived from the second derivative $d^{2}V/dI^{2}$ at 13 K. Thus, these TLF's are not due to phonons,~\cite{Akimenko1984,Matej1993} but could be due to rearrangements of cluster of atoms in both sides of the PC.~\cite{Ralls1991,Holweg1992prb}

For PC `F' that requires a large depairing parameter $\Gamma$ to fit the PCAR spectrum, TLF's strongly dominate the noise spectra as shown in Fig.~4. The normalized $fS_{V}/V^{2}$ at different frequencies do not coincide with each other due to discrete TLF's in the whole bias range. The position and magnitude of the TLF's changes dramatically during repeated measurements at the same temperature and at higher temperatures, which suggests that these TLF's can be easily heated up and get rearranged, even though the PC itself is still ballistic. Owing to this fact we argue that the large $\Gamma$ might arise from the TLF's in the PC `F'. We believe that even in the point-contacts between normal metals and unconventional superconductors, the large $\Gamma$ might originate from such TLF's and this aspect should be considered while analyzing PCAR data with the BTK theory.

This research was conducted with support from U.S. Department of Energy through Grant No. DE-FG02-06ER46346.

\end{document}